\documentclass[twocolumn,english,notitlepage]{revtex4-1}
\usepackage[T1]{fontenc}
\usepackage[utf8]{inputenc}
\usepackage{color}
\definecolor{shadecolor}{rgb}{1, 0, 0}
\usepackage{verbatim}
\usepackage{framed}
\usepackage{amsmath}
\usepackage{graphicx}
\usepackage[unicode=true]
 {hyperref}
\usepackage{breakurl}

\makeatletter
\usepackage{breqn}
\usepackage{enumitem}
\usepackage{babel}

\makeatother

\begin{document}

\title{LongHCPulse: Long Pulse Heat Capacity on a Quantum Design PPMS}

\author{Allen Scheie}

\affiliation{Institute for Quantum Matter and Department of Physics and Astronomy
Johns Hopkins University Baltimore MD 21218}

\date{\today}
\begin{abstract}
This paper presents LongHCPulse: software which enables heat capacity
to be collected on a Quantum Design PPMS using a long-pulse method.
This method, wherein heat capacity is computed from the time derivative
of sample temperature over long (30 min) measurement times, is necessary
for probing first order transitions and shortens the measurement time
by a factor of five. LongHCPulse also includes plotting utilities
based on the Matplotlib library. I illustrate the use of LongHCPulse
with the example of data taken on ${\rm Yb_{2}Ti_{2}O_{7}}$, and
compare the results to the standard semi-adiabatic method.
\end{abstract}

\maketitle

\section{Introduction}

Low temperature heat capacity is an important measurement in materials
characterization. It is used to study phase transitions \cite{Tari},
nuclear magnetism \cite{Mirebeau_Tb2Sn2O7_schottky}, superconductivity
\cite{Kemper2016}, and magnetic properties \cite{Applegate2012,LiCuSbO_2012,Yamashita2011,Pomaranski2013}.
A common tool for measuring low temperature heat capacity is a Quantum
Design Physical Properties Measurement System (PPMS) \cite{PPMS_Manual},
which can measure heat capacity from 400~K down to 50~mK with a
dilution refrigerator insert. Unfortunately, low-temperature measurements
are time-consuming, and the standard PPMS measurement technique is
not sensitive to the latent heat of first order phase transitions.
We present here a program for accurately measuring heat capacity using
a PPMS with a long-pulse method that is sensitive to first-order transitions
and decreases measurement time by more than a factor of five. This
software enables a new measurement technique on a common piece of
lab equipment. 

By default, the PPMS measures heat capacity with a semi-adiabatic
thermal relaxation technique \cite{Shi20101107,PPMS_Manual}, wherein
a short heat pulse is applied to the sample, and heat capacity is
extracted from exponential curve fits to the heating and cooling data.
(I refer to this as the ``semi-adiabatic method'' in this paper.)
Useful as it is, this approach has two limitations. First, the semi-adiabatic
method assumes constant heat capacity over the entire pulse range,
rendering this method unsuitable for probing first-order phase transitions
\cite{Lashley2003369}. Secondly, the semi-adiabatic method can be
very time-consuming when measuring multiple magnetic fields. A typical
constant-field PPMS heat capacity measurement between 2~K and 0.1~K
can take 30 hours, meaning that mapping the $H-T$ plane with 20 magnetic
fields can take as long as a month. 

An alternative to the semi-adiabatic method is a ``long-pulse method'',
wherein one applies a long (\textasciitilde{}30 min) heat pulse to
the sample (causing a temperature rise of up to 200\%) and computes
heat capacity from the time derivative of sample temperature. This
technique is accurately sensitive to first-order transitions because
data is collected as the sample continuously passes through the full
temperature range (unlike the method used in Ref. \cite{Lashley2003369})
and saves a significant amount of time. The PPMS MultiVu software
has a dual-slope analys feature which is meant to do this, but the
MultiVu software loses accuracy and consistency when the heat pulses
cover a significant temperature range. 
To properly handle long-pulse heat capacity data, I present LongHCPulse,
a Python software package which enables efficient processing of long-pulse
data collected with a Quantum Design PPMS from 40~K down to dilution
refrigerator (DR) temperatures. 

This software was developed in order to process and plot heat capacity
data for ${\rm Yb_{2}Ti_{2}O_{7}}$ \cite{Scheie_2017}. The data
shown in this paper is from this compound.

\section{Background}

\subsection{Theoretical Background: Heat Capacity from Slope Analysis}

Heat capacity can be computed from the time derivative of sample temperature
if the heat flow into and out of the sample is known. The PPMS sample
stage is a good setup for this method.

\begin{figure}
\centering\includegraphics[scale=0.5]{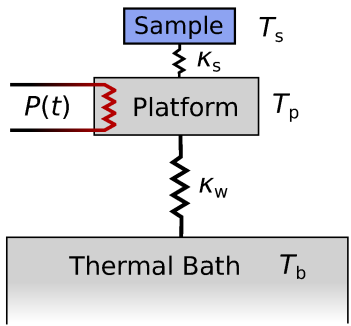}

\caption{Schematic diagram of the PPMS dilution refrigerator sample stage.
Usually $\kappa_{s}\gg\kappa_{w}$, and so $T_{s}=T_{p}$.}
 \label{flo:StageDiagram}
\end{figure}

A schematic diagram of a PPMS dilution refrigerator measurement stage
(also called a heat capacity puck) is shown in Fig. \ref{flo:StageDiagram}.
The sample is thermally connected to a platform with Apiezon N grease,
and the platform is thermally coupled to a temperature bath at temperature
$T_{b}$ with a wire of known thermal conductivity $\kappa_{w}$.
Embedded within the sample platform lies a small resistive heater
which applies a well-controlled power $P(t)$. For measurements above
80~mK, the grease has a negligibly small thermal resistivity such
that both the sample and the platform are at the same temperature
$T_{s}$. Using these variables, we can derive an expression for sample
heat capacity.

We begin with the equations for heat capacity $Q=C\frac{dT_{s}}{dt}+P(t)$
and heat flow through the wire $\frac{dQ}{dx}=\kappa_{w}\frac{dT}{dx}$.
$Q$ is net heat flow into the sample, $C$ is sample heat capacity,
$T_{s}$ is the sample temperature, $P(t)$ is power applied to the
sample platform, and $\kappa_{w}$ is the wire conductivity (measured
during puck calibration). Using the second equation, we can integrate
over $dx$ to obtain the heat flow $Q$ through the wire:
\begin{equation}
Q=\int_{T_{b}}^{T_{s}}\kappa_{w}~dT\label{eq:hc3}
\end{equation}
In general, $\kappa_{w}$ depends on temperature, so this numerical
integral needs to be carried out explicitly. Combining the above equations,
we arrive at an expression for heat capacity:
\begin{equation}
C=\left(\frac{dT_{s}}{dt}\right)^{-1}\int_{T_{b}}^{T_{s}}\kappa_{w}\,dT-P(t)\label{eq:HC}
\end{equation}

In practice, there are other sources of heat loss such as thermal
radiation which are not included in this equation. To account for
this, the PPMS software manual recommends adding an ad-hoc correction
factor to $\kappa_{w}$ \cite{PPMS_Manual} making the final equation
\begin{equation}
C=\left(\frac{dT_{s}}{dt}\right)^{-1}\int_{T_{b}}^{T_{s}}(\kappa_{w}(T)+\kappa_{os})dT-P(t)\label{eq:HC_corrected}
\end{equation}
where $\kappa_{os}$ is $\kappa_{w}(T_{b})\times S$, where $S$ is
a phenomenologically determined static offset parameter. Some radiative
heat loss is always present, and a small offset of $S=0.1$ was necessary
to make our measurements self-consistent.

The measurement .raw file and the calibration .cal files contain all
the information necessary to compute heat capacity using Eq. \ref{eq:HC_corrected}.

\subsection{PPMS Software}

\begin{figure}
\centering\includegraphics[scale=0.58]{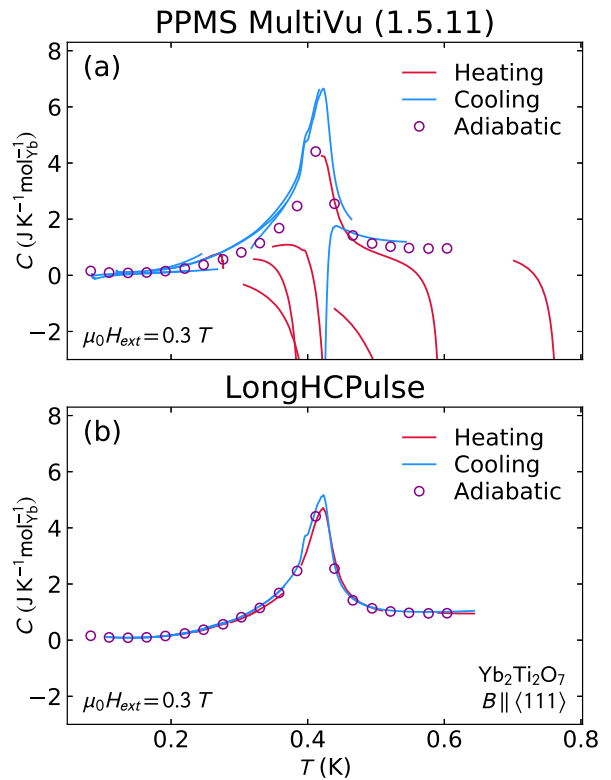}

\caption{Comparison of long-pulse slope analysis heat capacity computed using
the PPMS software and LongHCPulse. (a) Heat capacity computed using
the PPMS MultiVu software (moving average $n=5$). The heating pulses
are wildly inconsistent and give unphysical values. The cooling pulses
are somewhat self-consistent, but not consistent with the semi-adiabatic
method data. (b) Heat capacity computed using LongHCPulse. The heating,
cooling, and semi-adiabatic data are all self-consistent, though there
is a slight difference between heating and cooling pulses. The semi-adiabatic
data was collected by starting at high temperature and going to low
temperature.}

\label{flo:PPMSvsLongHCPulse}
\end{figure}

In principle, the PPMS MultiVu software can carry out the calculations
in Eq.~\ref{eq:HC_corrected} using its ``Slope Analysis'' feature.
However, our version of the PPMS software (MultiVu 1.5.11) generates
inconsistent and incorrect heat capacity values when the heat pulses
cover a significant temperature range. Large temperature ranges are
necessary for probing first order transitions, because the hysteresis
between heating and cooling can only be accurately recorded if a pulse
traverses the entire transition. As Fig.~\ref{flo:PPMSvsLongHCPulse}(a)
shows, the long-pulse data processed with the PPMS software is neither
self-consistent nor physically sensible (negative heat capacity values
are impossible). Clearly, a different approach is needed. 

\section{Results: LongHCPulse Code}

LongHCPulse is a python class which computes and plots heat capacity
from PPMS long-pulse data. As Fig.~\ref{flo:PPMSvsLongHCPulse}(b)
shows, heat capacity computed with LongHCPulse is both self-consistent
and in agreement with semi-adiabatic short-pulse data. 

The implementation as a Python class allows for easy implementation
in Python scripts; heat capacity can be computed in as few as five
lines of code. In addition, all data is stored in instance variables,
giving the user freedom to access or modify the data as desired. Finally,
the LongHCPulse package uses the popular Matplotlib library to provide
post-processing and plotting utilities.

\subsection{Inputs}

LongHCPulse requires the following information: the .raw data file
from the PPMS measurement, the calibration file for the sample puck
used in the measurement, and the sample mass and molar mass.

\subsection{Automatic Corrections}

To properly compute heat capacity, LongHCPulse makes three automatic
corrections to the raw data, two of which are made in the MultiVu
software, and one of which is not.

First, LongHCPulse smooths the temperature data before taking the
derivative $\frac{dT_{s}}{dt}$, as in the PPMS MultiVu software.
A derivative of noisy data is almost unintelligible, so a simple moving
average (default $n=5$) is applied to the data.

Second, LongHCPulse eliminates the data from the beginnings and ends
of the pulses (default within 15\% of maximum or minimum temperature).
The data at the very beginning of a pulse is unreliable because the
thermal equilibration between the sample and sample platform is not
instantaneous due to finite $\kappa_{s}$ (see Fig. \ref{flo:StageDiagram}).
Consequently, the initial temperature readings do not reflect actual
the sample temperature. LongHCPulse corrects for this by ignoring
the first few data points after the heater turns on or off. Data at
the end of the pulse are eliminated because there $\frac{dT_{s}}{dt}$
becomes very small, rendering the uncertainty very large (see Appendix
\ref{sec:Uncertainty-for-Heat}) and the data unreliable. By default,
LongHCPulse eliminates data within 15\% of the maximum or minimum
temperatures. (The user may choose to modify this.) This step is also
possible with the PPMS MultiVu software's slope analysis feature \cite{PPMS_Manual}.

\begin{figure}
\centering\includegraphics[scale=0.45]{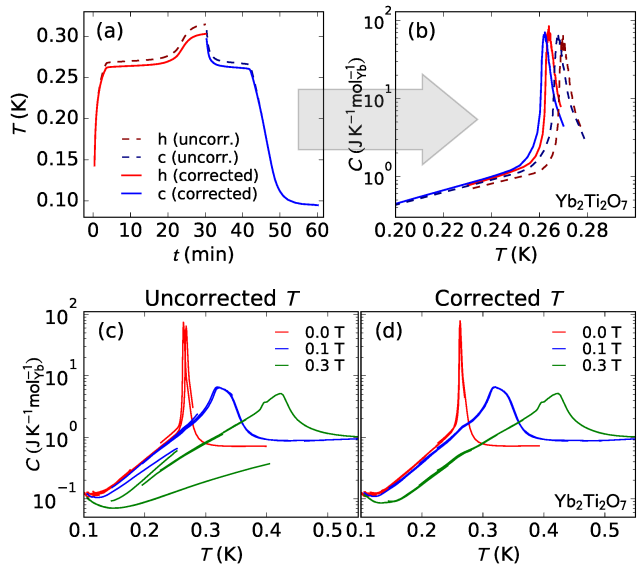}

\caption{Effects of correcting the temperature values by re-computing from
thermometer resistance. (a) Heating and cooling pulses for a zero-field
${\rm Yb_{2}Ti_{2}O_{7}}$ measurement, corrected and uncorrected.
Note the plateau at 0.27~mK due to the first order phase transition.
(b) Heat capacity computed from the data in panel (a). (c) Heat capacity
of ${\rm Yb_{2}Ti_{2}O_{7}}$ computed from multiple pulses in multiple
fields without correcting the temperature (cooling pulses only, to
avoid confusion from the hysteresis in zero-field data). Note the
discrepancy between different measurements. (d) Heat capacity computed
from the same pulses as (c), correcting the temperature. }
\label{flo:TemperatureCorrection}
\end{figure}

Third, LongHCPulse recomputes every temperature value from the thermometer
resistance. When $(T_{b}-T_{s})$ becomes significant, heat capacity
from different heat pulses are inconsistent if one uses $T_{s}$ values
recorded in the raw data file (see Fig. \ref{flo:TemperatureCorrection}).
However, if one re-computes $T_{s}$ from thermometer resistance (using
the $T$ vs $R$ curves in the DR puck calibration file), the resulting
heat capacity data are self-consistent. This inaccuracy is possibly
due to the PPMS software using only a small interpolation range, leading
to errors as $(T_{b}-T_{s})$ becomes large. In any case, all temperature
values need to be re-computed from thermometer resistance.

All these steps are automatically taken in 'LongHCPulse'.

\subsection{Manual Correction: Wire Conductivity}

For quantitatively accurate results, it is necessary to measure some
short-pulse semi-adiabatic data at at least one magnetic field (which
is a good practice anyway) to obtain thermal conductivity values for
the specific measurement.

Although the wire conductivity $\kappa_{w}$ is measured in the puck
calibration, it varies slightly from measurement to measurement \cite{kennedy2007recommendations}.
This is because the sample shape, position on the platform, and internal
conductivity all influence the effective value of $\kappa_{w}$. (For
instance, a thick sample with a low internal conductivity will have
a much higher effective $\kappa_{w}$ than a thin sample with high
internal conductivity.) Therefore, it is necessary to correct $\kappa_{w}$
for every measurement. Fortunately, this is easy to do by taking some
semi-adiabatic data. 

\begin{figure}
\centering\includegraphics[scale=0.48]{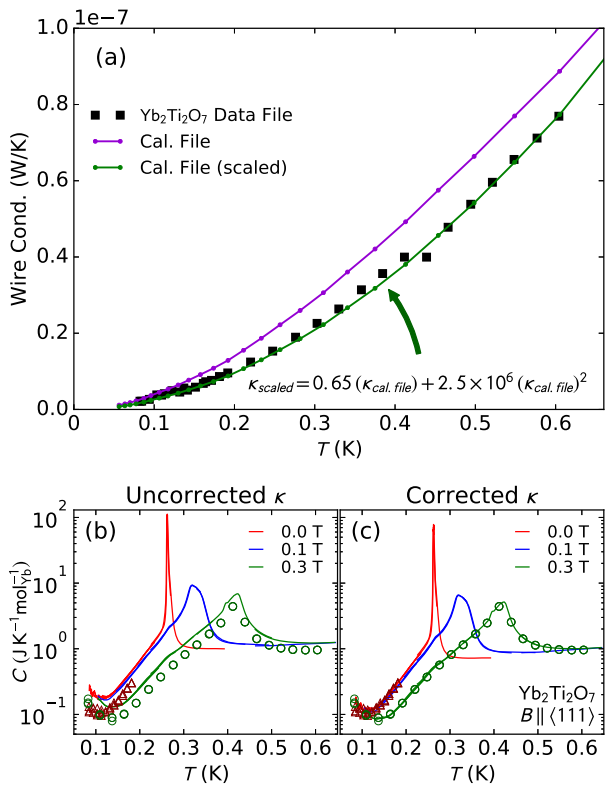}

\caption{(a) Thermal conductivity of the wire connecting the sample stage to
the thermal bath. Note the difference between the calibration measurement
and the values extracted from the short-pulse data, necessitating
the correction. (b) Heat capacity computed from the uncorrected ``Cal.
File'' thermal conductivity (cooling pulses only). Note the discrepancy
between the long-pulse and short-pulse data. (c) Heat capacity computed
with the corrected ``Cal. File (scaled)'' thermal conductivity values.}

\label{flo:ThermCondCorrection}
\end{figure}

When the PPMS measures heat capacity with the semi-adiabatic method,
the wire conductivity is among the parameters extracted from the fit.
This provides an appropriate reference with which one can correct
$\kappa_{w}$. Thus, the user must scale \texttt{self.Kw} manually
before computing heat capacity; LongHCPulse does not make this correction
automatically. (Directly using the thermal conductivity values from
the short-pulse measurements can yield very noisy results.) Typically,
a linear plus a quadratic correction term suffices. As Fig. \ref{flo:ThermCondCorrection}
shows, this correction makes the long-pulse data match the semi-adiabatic
data—though the general shape of the heat capacity peaks are still
accurate without this correction.

\subsection{Features}

In addition to computing heat capacity from raw data files, LongHCPulse
contains many features which streamline data processing: 
\begin{itemize}
\item If semi-adiabatic short pulse data has been taken in as a part of
the same measurement, LongHCPulse will recognize short pulses and
treat them and plot them separately.
\item The heat capacity data can be scaled by an arbitrary factor with \texttt{LongHCPulse.scale()}.
The data output by LongHCPulse is, by default, in units of ${\rm \frac{J}{K\,mol_{F.U.}}}$.
A separate scaling factor, applied during initialization, is required
for the semi-adiabatic short-pulse data, which is typically recorded
in $\mu{\rm J/K}$.
\item All magnetic fields can be scaled by a demagnetization factor with
\texttt{LongHCPulse.scaleDemagFactor()}. (This correction is often
necessary when examining low-field behavior.) To do this, the user
must provide LongHCPulse with a 2D numpy array with $M$ vs $H_{int}$
in units of ${\rm \frac{A}{m}}$ and ${\rm Oe}$, respectively.
\item All processed data can be saved to a .pickle file. This can then be
imported in a separate Python script by instantiating the class with
the saved file. This is useful because (i) the same data can be plotted
and manipulated in multiple scripts without duplicating code, and
(ii) the data processing itself can take 1-2 minutes. Processing the
data in one script and plotting the data in another script saves time
when making small adjustments to plots.
\end{itemize}

\subsubsection*{Plotting Utilities}

LongHCPulse has many utilities for plotting heat capacity data using
the Matplotlib library \cite{MatPlotLib}:
\begin{itemize}
\item Individual heating and cooling pulses can be plotted using \texttt{LongHCPulse.plotHC()}.
$C/T$ can be plotted with \texttt{LongHCPulse.plotHCT()}. The user
has the option of displaying any combination of heating pulses, cooling
pulses, and semi-adiabatic data.
\item \texttt{LongHCPulse.lineplot()} plots all data taken within 10~Oe
of a specified magnetic field. If multiple fields are specified, then
the data from each field is plotted with a different color from a
rainbow colormap.
\item \texttt{LongHCPulse.lineplotCombine()} combines all data within 10~Oe
of a specified magnetic field into a single trace. In the regions
where two or more traces overlap in temperature, the average heat
capacity value is computed. If multiple fields are specified, each
field is plotted with a different color from a rainbow colormap, just
like \texttt{lineplot(). }Bin size may be increased by modifying the
FieldBinSize function variable.
\item \texttt{LongHCPulse.plotEntropy()} computes the entropy change $\Delta S=\int\frac{C}{T}dT$
for every magnetic field. It first combines the data in the same manner
as \texttt{lineplotCombine()}, and plots them with the same rainbow
colormap scheme.
\item \texttt{LongHCPulse.meshgrid()} creates a 3D array of the data ($C$
vs $T$ and $B$) that can be plotted with \texttt{matplotlib.axes.Axes.pcolormesh()}.
\end{itemize}
For all plotting options, the user may selectively plot heating and
cooling pulses.

\subsection{Guidelines for Collecting Data}

The following are some guidelines for collecting long-pulse data:

\paragraph{MultiVu Commands:}

When collecting long-pulse data on a Quantum Design PPMS, one must
use a special sequence of software commands. A sample set of commands
for one pulse is:

{ \definecolor{shadecolor}{rgb}{0.85,0.85,0.85}

\begin{shaded}%
Set Temperature 0.22K at 1K/min, Fast Settle

Set Magnetic Field 800.0Oe at 100.0Oe/sec, Linear, Persistent 

Wait For Temperature, Field, Delay 60 secs, No Action

Sample HC at current temperature, 0.300 K rise, 1 times, 3000 sec
meas time, simple fit, no settling\end{shaded}

} This will be repeated for each base temperature and each magnetic
field.

\paragraph{Pulse Length:}

The length of the pulse (``meas time'') depends upon the sample
being measured, and must be optimized by the user. Pulses that are
too short will not reach the full temperature range, but pulses that
are too long will waste time by allowing the sample to sit at the
maximum temperature. The first few measurements should be used to
optimize the ``meas time'' parameter. For a 1.04~mg ${\rm Yb_{2}Ti_{2}O_{7}}$
sample, long measuring times (3000 s) were required for low-fields
to overcome the latent heat of the transition, but only 800 s were
required to cover the same temperature range at higher fields (where
the peak in heat capacity was diminished).

\paragraph{Pulse Size:}

This program was designed to process data taken below 1~K for pulses
covering 200-300~mK. However, preliminary tests show that LongHCPulse
works at temperatures up to 120~K for pulses up to 30~K, but is
not reliably accurate for pulses above 40K (see Appendix \ref{sec:Temperature-Range}).
Typically, it is better to measure heat capacity at different temperatures
with overlapping temperature ranges than repeat long-pulses at the
same base temperature. Also, it saves time to to loop through magnetic
fields at a constant base temperature: repeatedly changing the base
temperature is far slower than changing the magnetic field.

\section{Discussion: Software Performance}

\subsection{Accuracy}

Two measures of accuracy for LongHCPulse are (i) self-consistency
between pulses taken at different base temperatures and (ii) agreement
with the semi-adiabatic method. As Fig. \ref{flo:PPMSvsLongHCPulse}
shows, LongHCPulse scores well on both of these measures.

\subsection{Data Acquisition Speed}

It took a total of 105 hours to collect the ${\rm Yb_{2}Ti_{2}O_{7}}$
long-pulse heat capacity data in Ref. \cite{Scheie_2017} covering
21 magnetic fields (not including initial cool-down, but including
optimization of meas time). In the same experiment, it took 42 hours
to measure 1.5 temperature sweeps using the semi-adiabatic method.
This is a speed-up of at least a factor of five. Subsequent tests
showed that 40\% more time may be saved if the user is judicious with
taking data.

\subsection{Uncertainty}

One can estimate uncertainty in long-pulse heat capacity using the
general formula for statistical uncertainty (see Appendix \ref{sec:Uncertainty-for-Heat}).
This approach has the weakness of ignoring correlated uncertainties,
and therfore is meant to give an approximate understanding. 

\begin{figure}
\centering\includegraphics[scale=0.44]{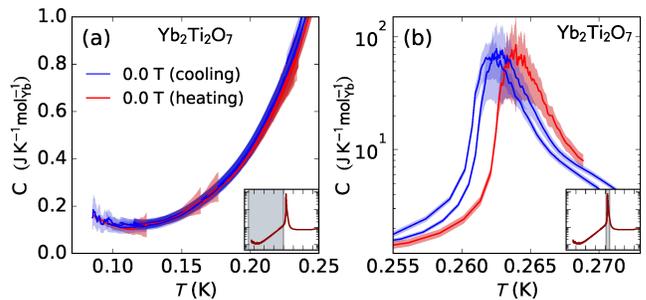}

\caption{Uncertainty of zero-field ${\rm Yb_{2}Ti_{2}O_{7}}$ heat capacity
in two different temperature ranges (temperature ranges shown by insets).
Lightly shaded regions indicate calculated error bars.}
\label{flo:uncertainty}
\end{figure}

According to Eq. \ref{eq:uncertainty} in Appendix \ref{sec:Uncertainty-for-Heat},
$\delta C$ is inversely proportional to $\frac{dT_{s}}{dt}$. This
means that the uncertainty becomes very large when $T_{s}$ is nearly
constant, like when the system goes through a first-order phase transition
or reaches the end of a pulse. To visualize uncertainty, I computed
uncertainty assuming $\delta T_{s}=0.03\,{\rm mK}$, $\delta T_{b}=0.1\,{\rm mK}$,
$\delta P=10^{-13}{\rm W}$, $\delta S=0.01$, and computing $\delta\kappa_{w}$
from the standard deviation of $\kappa_{w}$ values extracted from
semi-adiabatic measurements. Fig. \ref{flo:uncertainty} shows the
calculated uncertainty. Note how the uncertainty is larger at the
top of the heat capacity peak and at the end of pulses. Away from
the heat capacity peak, the variation in $C/T$ between pulses falls
within the error bars. On top of the peak, the computed heat capacity
does not agree to within uncertainty. I attribute this to slight thermal
hysteresis coming from some of the pulses not making it all the way
over the first order transition peak—an effect which Eq. \ref{eq:uncertainty}
does not account for.

If the user desires to see the error bars computed with the above
parameters, add the command ``PlotUncertainty=True'' to any of the
LongHCPulse plot commands, and error bars will be plotted like in
Fig. \ref{flo:uncertainty}. This feature is not available for combined
traces.

\section{Example: ${\rm Yb_{2}Ti_{2}O_{7}}$}

The code for LongHCPulse is available at \href{https://github.com/asche1/LongHCPulse}{https://github.com/asche1/LongHCPulse},
which also contains the notebook used to process the data in Ref.
\cite{Scheie_2017}. This repository will be continually updated with
the most recent version of LongHCPulse. Please report bugs to scheie@jhu.edu.

\section{Conclusion}

LongHCPulse is a software package which allows a PPMS to measure heat
capacity with a long-pulse method, and yields results that are both
self-consistent and in agreement with the semi-adiabatic method. This
method is sensitive to first-order transitions, and it saves a significant
amount of time in collecting data. The implementation in Python allows
for flexibility to process data and easy assimilation into Matplotlib
scripts. Once incorporated into a researcher's workflow, LongHCPulse
can save time and eliminate human error by automating routine analysis.
With this software, the capabilities of the Quantum Design PPMS are
expanded, and heat capacity can be accurately measured with a new
technique.

\section*{Acknowledgments}

Thanks to Kate Arpino, James R. Neilson, and Zachary Kelley for allowing
me to test LongHCPulse with their data. Thanks also to Tyrel M. McQueen
for additional helpful suggestions. The author was supported through
the Gordon and Betty Moore foundation under the EPIQS program GBMF4532.

\newpage{}

\appendix

\section{\label{sec:Uncertainty-for-Heat}Uncertainty}

{\small{}We begin with Eq. \ref{eq:HC_corrected} written out explicitly
for a single data point
\[
C_{i}=\frac{\int_{T_{b_{i}}}^{T_{s_{i}}}\kappa_{w}(T)dT+S\,\kappa(T_{b})(T_{s_{i}}-T_{b_{i}})-P_{i}}{\frac{8(T_{s_{i+1}}-T_{s_{i-1}})-(T_{s_{i+2}}-T_{s_{i-2}})}{12(t_{i+1}-t_{i})}},
\]
where the denominator is the second-order finite central difference
derivative. The general formula for uncertainty is
\begin{equation}
\delta F=\sqrt{\left(\frac{\partial F}{\partial x_{1}}\delta x_{1}\right)^{2}+\left(\frac{\partial F}{\partial x_{2}}\delta x_{2}\right)^{2}+...}\quad.
\end{equation}
}{\small \par}

{\small{}The variables in our equation for heat capacity point $C_{i}$
are $T_{s_{i}}$, $T_{b_{i}}$, $\kappa_{w}$, $P_{i}$, $S$, $T_{s_{i+1}}$,
$T_{s_{i-1}}$, $T_{s_{i+2}}$, $T_{s_{i-2}}$, the last four being
for the numerical derivative of $\frac{dT_{i}}{dt}$. We write out
the derivatives with respect to all of these:
\begin{align*}
\frac{\partial C_{i}}{\partial T_{s_{i}}} & =\,\frac{1}{\frac{dT_{s_{i}}}{dt}}\frac{\partial}{\partial T_{s_{i}}}\left(\int_{T_{b_{i}}}^{T_{s_{i}}}\kappa_{w}(T)dT+S\,\kappa_{w}(T_{b_{i}})(T_{s_{i}}-T_{b_{i}})\right)\\
 & =\left(\frac{dT_{s_{i}}}{dt}\right)^{-1}\left(\kappa_{w}(T_{s_{i}})+S\,\kappa_{w}(T_{b_{i}})\right)
\end{align*}
}
{\small \par}

{\small{}similarly,
\[
\frac{\partial C_{i}}{\partial T_{b_{i}}}\,=\,\left(\frac{dT_{s_{i}}}{dt}\right)^{-1}(1-S)\kappa_{w}(T_{b_{i}})
\]
}{\small \par}

{\small{}
\begin{align*}
\frac{\partial C_{i}}{\partial\kappa} & =\,\left(\frac{dT_{s_{i}}}{dt}\right)^{-1}\frac{\partial}{\partial\kappa}\left(\int_{T_{b_{i}}}^{T_{s_{i}}}\kappa_{w}(T)dT+S\,\kappa_{w}(T_{b_{i}})(T_{s_{i}}-T_{b_{i}})\right)\\
 & =\left(\frac{dT_{s_{i}}}{dt}\right)^{-1}(1+S)(T_{s_{i}}-T_{b_{i}})
\end{align*}
\[
\frac{\partial C_{i}}{\partial P_{i}}\,=-\left(\frac{dT_{s_{i}}}{dt}\right)^{-1}.
\]
\[
\frac{\partial C_{i}}{\partial S}\,=\,\left(\frac{dT_{s_{i}}}{dt}\right)^{-1}\kappa_{w}(T_{b_{i}})(T_{s_{i}}-T_{b_{i}})
\]
\[
\frac{\partial C_{i}}{\partial T_{s_{i\pm1}}}=\mp C_{i}\left(\frac{dT_{s_{i}}}{dt}\right)^{-1}\frac{8}{12(t_{i}-t_{i+1})}
\]
}{\small \par}

{\small{}}{\small \par}

{\small{}similarly,}{\small \par}

{\small{}
\[
\frac{\partial C_{i}}{\partial T_{s_{i\pm2}}}\,=\mp C_{i}\left(\frac{dT_{s_{i}}}{dt}\right)^{-1}\frac{1}{12\Delta t}.
\]
}{\small \par}

{\small{}Assuming $\delta T_{s_{i}}\approx\delta T_{s_{i\pm1}}\approx\delta T_{s_{i\pm2}}$,
our general formula for uncertainty is:
\begin{align}
\delta C_{i} & =\left(\frac{dT_{s_{i}}}{dt}\right)^{-1}\Bigg[\left((1-S)\kappa_{w}(T_{b_{i}})\delta T_{b_{i}}\right)^{2}+\delta P_{i}^{2}\nonumber \\
 & +\left(\kappa_{w}(T_{b_{i}})(T_{s_{i}}-T_{b_{i}})\delta S\right)^{2}+\left((1+S)(T_{s_{i}}-T_{b_{i}})\delta\kappa\right)^{2}\nonumber \\
 & +\left(\left(\kappa_{w}(T_{s_{i}})+S\,\kappa_{w}(T_{b_{i}})\right)^{2}+\frac{65\,C_{i}^{2}}{72\,\Delta t^{2}}\right)\delta T_{s_{i}}^{2}\Bigg]^{\frac{1}{2}}\label{eq:uncertainty}
\end{align}
}As noted in the text, this equation ignores correlated uncrtainties.
For example, $\delta T_{s_{i+1}}$ is correlated with $\delta T_{s_{i-1}}$,
$\delta T_{s_{i+2}}$ and $\delta T_{s_{i-2}}$ because they are calculated
from the same $R$ vs $T$ curve, causing any propogated errors to
shift all $T_{s}$ readings the same direction. Similarly, $\delta T_{s}$
is correlated to $\delta T_{b}$ because they both come from the same
thermometer. Because of this, the uncertainties shown here ought to
be taken with a grain of salt; they are meant merely to give an overiew.

\section{Temperature Range\label{sec:Temperature-Range}}

\begin{figure}
\centering\includegraphics[scale=0.4]{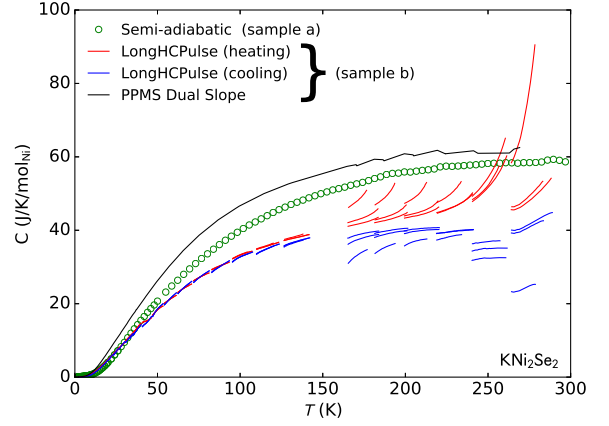}

\caption{Temperature range of LongHCPulse shown by ${\rm KNi_{2}Se_{2}}$ heat
capacity data. LongHCPulse is accurate below 40~K, but deviates from
the semi-adiabatic data above 40~K. Above 120~K, LongHCPulse breaks
down and the heating pulses fail to match the cooling pulses.}
\label{flo:KNiSe}

\end{figure}

The temperature range of LongHCPulse was tested with old data taken
on ${\rm KNi_{2}Se_{2}}$ (see Fig. \ref{flo:KNiSe}). Unfortunately,
the semi-adiabatic data were acquired on a different sample, so the
$\kappa_{w}$ values (and thus the heat capacity values) are probably
a little off. Nonetheless, it is possible to get a rough picture of
the temperature range of LongHCPulse. As currently implemented, LongHCPulse
matches semi-adiabatic data below 40K, and retains qualitative accuracy
up to 120~K. Above 120~K, additional temperature gradients are probably
present throughout the sample chamber, causing variation in both $\kappa_{w}$
and $T_{b}$ and wild inconsistency in heat capacity values.

\section{Pseudocode}

The Pseudocode for the main functions in LongHCPulse is as follows:

{ \definecolor{shadecolor}{rgb}{0.85,0.85,0.85}

\begin{shaded}%
\noindent \_\_init\_\_(data\_file, calibration\_file, sample\_mass,
sample\_molar\_mass, short-pulse\_scale\_factor):
\begin{itemize}
\item [] \setlength{\itemsep}{0pt} \setlength{\parskip}{-2pt} Import calibration
file:
\begin{itemize}
\item [] \setlength{\itemsep}{-1mm} Import $\kappa_{w}$ vs. $T$
\item [] Import thermometer resistance vs. $T$ and $B$
\end{itemize}
\item [] Import raw data:
\begin{itemize}
\item [] \setlength{\itemsep}{0mm} FOR pulses:
\begin{itemize}
\item [-] Collect time, heater power, and resistance
\item [-] Re-compute temperature from thermometer resistance using calibration
data
\item [-] Collect info about pulse (field, $T_{b}$, addenda, etc.)
\item [-] IF $(T_{max}-T_{min})/(T_{avg})<0.1$:
\begin{itemize}
\item [] Flag pulse as short pulse
\end{itemize}
\end{itemize}
\end{itemize}
\item [] FOR ShortPulses:
\begin{itemize}
\item [] Average all short pulses within 1\% of $T_{sample}$
\end{itemize}
\item [] Take moving average of temperature data (n=5)
\end{itemize}
~

\noindent Compute Heat Capacity()
\begin{itemize}
\item [] \setlength{\itemsep}{0pt} \setlength{\parskip}{-2pt} FOR Pulses:
\begin{itemize}
\item [] \setlength{\itemsep}{0pt} FOR DataPoints:
\begin{itemize}
\item [-] Compute $dT/dt$ using 2nd order central finite difference method
\item [-] Compute heat flow through wires with $\int_{T_{b}}^{T_{s}}\kappa_{w}~dT$
\item [-] Compute heat capacity ($C$) from above values
\item [-] Subtract addenda heat capacity
\item [-] Compute uncertainty
\item [-] Eliminate $C$ values too close to $T_{min}$ or $T_{max}$ 
\end{itemize}
\end{itemize}
\item [] Convert to J/K/mol from J/K
\end{itemize}
~	

\noindent Scale for Demagnetization Factor($D$)
\begin{itemize}
\item []\setlength{\itemsep}{0pt} \setlength{\parskip}{-2pt} FOR all magnetic
fields: 
\begin{itemize}
\item []\setlength{\itemsep}{-1pt} Solve $H_{int}=H_{ext}-D\times M(H_{int})$
for $H_{int}$ using the bisection method
\end{itemize}
\end{itemize}
~

\noindent Line Plot Combine(B fields)
\begin{itemize}
\item [] \setlength{\itemsep}{0pt} \setlength{\parskip}{-2pt} FOR specified
B fields:
\begin{itemize}
\item [-] \setlength{\itemsep}{-1pt} Identify traces which overlap in temperature
and compute average of data points for the overlap region
\item [-] Concatenate cooling pulse data into single array
\item [-] Sort array from low $T$ to high $T$
\end{itemize}
\item [] Plot data from specified fields with line plots
\end{itemize}
~

\noindent Meshgrid(Tarray, Barray):
\begin{itemize}
\item []\setlength{\itemsep}{0pt} \setlength{\parskip}{-2pt} Bin long-pulse
data into 2D histogram
\item [] return 3D numpy array suitable for plt.pcolormesh()
\end{itemize}
\end{shaded}

}
\end{document}